\documentstyle[sprocl,epsfig]{article}

\input{psfig.sty}

\def\One{1\kern-4.5pt1}

\def\be{\begin{equation}}
\def\ee{\end{equation}}
\def\bea{\begin{eqnarray}}
\def\eea{\end{eqnarray}}
  \newcommand{\epsfaxhax}[2]{
          \centerline{
            \hspace{-15pt}
            \epsfxsize=160pt
            {\epsfbox{#1}}
            \hspace{-7pt}
            \epsfxsize=160pt
            {\epsfbox{#2}}}
  }

\begin{document}

\title{TWO COLOURS QCD AT NONZERO CHEMICAL POTENTIAL}

\author{Susan MORRISON and Simon HANDS}

\address{Department of Physics, University of Wales Swansea,\\
Singleton Park, Swansea, SA2 8PP, U.K.} 


\maketitle\abstracts{ We identify the Goldstone modes appropriate to
the low and high density phases of $SU(2)$ lattice gauge theory with
staggered fermions. We present hybrid Monte Carlo simulation results
for susceptibilities on a $6^4$ lattice at $\beta=1.5$, $m=0.05$. 
We specify how to implement a diquark source term in 
a lattice simulation and present first measurements of the lattice diquark 
condensate as a function of the diquark source.}

The symmetry breaking pattern of $SU(2)$ lattice gauge theory with 
fermions in the fundamental representation of the group is different 
from that of the continuum model where the symmetry breaking pattern 
for fermions in a pseudoreal representation is 
$SU(2N_{f})\rightarrow Sp(2N_{f})$ \cite{Peskin}. To identify the symmetries appropriate
to the lattice model with staggered fermions we start with the kinetic 
term in the lattice action
for a gauged isospinor doublet \cite{HT}
of staggered fermions. For clarity we consider $N=1$ flavors.

\bea
S_{kin}=& &\sum_{x,\,\nu=1,3}{{\eta_\nu(x)}\over2}\left[
\bar\chi(x)U_\nu(x)\chi(x+\hat\nu)-\bar\chi(x)U_\nu^\dagger(x-\hat\nu)
\chi(x-\hat\nu)\right]  \nonumber \\
&+ &\sum_{x}{{\eta_t(x)}\over2}\left[
\bar\chi(x)e^{\mu}U_{t}(x)\chi(x+\hat{t})-\bar\chi(x)e^{-\mu}U_{t}^\dagger(x-\hat{t})
\chi(x-\hat{t})\right].
\label{eq:sfund}
\eea
Defining:
\begin{equation}
\bar X_e=(\bar\chi_e,-\chi_e^{tr}\tau_2)\;\;:\;\;
X_o=\left(\matrix{\chi_o\cr -\tau_2\bar\chi_o^{tr}\cr}\right)
\end{equation}
and using $\eta_\mu(x\pm\hat\mu)=\eta_\mu(x)$ and 
$\tau_2U_\mu\tau_2=U_\mu^*$ (where $\tau_2$ is a Pauli matrix)
the kinetic term can be expressed in the following form: 
\begin{eqnarray}
S_{kin}=& &\sum_{x\;even,\,\nu=1,3}{{\eta_\nu(x)}\over2}\left[
\bar X_e(x)U_\nu(x)X_o(x+\hat\nu)-\bar X_e(x)U_\nu^\dagger(x-\hat\nu)
X_o(x-\hat\nu)\right] \nonumber \\
&+ &\sum_{x\;even}{{\eta_{t}(x)}\over2}\left[
\bar X_e(x)\left(\matrix{e^{\mu}&0\cr0&e^{-\mu}}\right) U_{t}(x)X_o(x+\hat{t})
- \right. \nonumber \\
& &\;\;\;\;\;\;\;\;\;\;\;\;\;\;\;\;\;\;\left.\bar X_e(x)\left(\matrix{e^{-\mu}&0\cr0&e^{\mu}}\right)U_{t}^\dagger(x-\hat{t})
X_o(x-\hat{t})\right]
\label{eq:sfundX}
\end{eqnarray}
There are two manifest global
$U(1)$ symmetries in the original action (\ref{eq:sfund}):

\be
\chi\mapsto e^{i\alpha}\chi\;\;\bar\chi\mapsto\bar\chi e^{-i\alpha}\;\;\;\;:
\;\;\;\chi\mapsto e^{i\alpha\varepsilon}\chi\;\;
\bar\chi\mapsto\bar\chi e^{i\alpha\varepsilon}
\ee
   where $\varepsilon(x)=(-1)^{x_1+x_2+x_3+x_4}$\,
The first invariance corresponds to baryon number conservation:
the second which holds only in the chiral limit ($m\rightarrow 0$) 
corresponds to conservation of axial charge.
Both of these are subsumed in a larger $U(2)$ symmetry which only holds
for $m=\mu=0$:
\begin{equation}
X_o\mapsto VX_o\;\;\bar X_e\mapsto\bar X_eV^\dagger\;\;\;\;V\in\mbox{U(2)}.
\label{eq:U2}
\end{equation}
In a continuum approach the axial anomaly would reduce the symmetry to 
$SU(2)$.

For $\mu\neq0$ the full $U(2)$ lattice symmetry is reduced to
$U(1)_{V}\otimes U(1)_{A}$ and is further reduced
to $ U(1)_{V}$ for $\mu\neq 0, m\neq 0$. It has been proposed 
\cite{Karsch86} that $U(1)_V$ is likely to be broken spontaneously by
the formation of a diquark condensate.
At $\mu=0$ the model displays {\sl spontaneous chiral symmetry
breaking}, with a chiral condensate of same form as the mass term
$\sum_x\bar\chi(x)\chi(x)$. 

\begin{equation}
\bar\chi\chi={1\over2}\left[\bar X_e\left(\matrix{&1\cr1&\cr}\right)
\tau_2\bar X_e^{tr}+
X_o^{tr}\left(\matrix{&1\cr1&\cr}\right)\tau_2X_o\right].
\label{eqn:qbq}
\end{equation}

This chiral condensate breaks the global symmetry (\ref{eq:U2}).
The residual symmetry is generated by the subgroup which leaves $\left(\matrix{0&1\cr1&0\cr}\right)$ invariant of which the
most general element is $\left(\matrix{e^{i\alpha}&\cr&e^{-i\alpha}\cr}
\right)$, which generates a $U(1)$. Thus we identify the pattern of
chiral symmetry breaking appropriate to the lattice model with
$N=1$ as $U(2)\to U(1)$. Now our hybrid Monte Carlo algorithm simulates 
with $\det M^{\dagger}M$ and since $\mbox{det}M^\dagger=\mbox{det}M^*=\mbox{det}\tau_2M\tau_2=\mbox{det}M$,
this describes two identical staggered fermion flavours resulting in
pattern of symmetry breaking $U(4) \to O(4)$, 
with 10 associated Goldstone modes. 

At high density, we postulate that a large Fermi surface will promote the formation of a diquark condensate. In principle many diquark 
wavefunctions can be written down therefore it is a dynamical question as to 
which condensate actually forms. We can deduce the ``maximally attractive
channel'' for the diquark condensate by insisting that the condensate 
wavefunction is anti-symmetric under exchange of fields \cite{ARW_S}
and by assuming the condensate is gauge invariant,
invariant under lattice parity and as local as possible in the $\chi$
fields. A local condensate $qq_{\bf2}$ which satisfies these conditions is:

\begin{equation}
qq_{\bf2}={1\over2}\left[\chi^{tr}(x)\tau_2\chi(x)+
\bar\chi(x)\tau_2\bar\chi^{tr}(x)\right].
\label{eq:qq2}
\end{equation}


In terms of the $X,\bar X$ fields the {\bf diquark condensate} is written
\begin{equation}
qq_{\bf2}={1\over2}\left[\bar X_e\left(\matrix{1&\cr&-1}\right)\tau_2
\bar X_e^{tr}+
X_o^{tr}\left(\matrix{1&\cr&-1}\right)\tau_2X_o\right].
\label{eq:dq2}
\end{equation}

 The diquark condensate (\ref{eq:dq2}) can be
obtained from the chiral condensate (\ref{eqn:qbq}) by an explicit global
U(2) rotation:
\begin{equation}
V={i\over{\surd{2}}}\left(\matrix{1&1\cr-1&1\cr}\right).
\end{equation}
Therefore in the limit $m=0$, $\mu=0$ the pattern of symmetry breaking remains $U(2) \to U(1)$: as
$\mu$ increases the condensate simply rotates from a chiral $\bar qq$ to 
a diquark $qq$. 

The full set of Goldstone modes (again for the  simplest case $N=1$)
can be found by considering infinitesimal rotations of either the chiral 
condensate (\ref{eqn:qbq}) in the low density phase or the diquark condensate
(\ref{eq:dq2}) in the high
density phase by
$V_\delta=\One+i\delta\lambda$, with $\lambda$ one of the $U(2)$ generators
$\{\One,\tau_i\}$, and identifying the mode as the coefficient of $O(\delta)$.
In the chiral limit and for $\mu=0$ we expect 3 massless Goldstone modes (since dim U(2)=4). 
Since the full $U(2)$ symmetry of the action is reduced for $\mu\neq 0$
only two of the four generators correspond to Goldstone modes i.e. 
the generators appropriate to $U(1)_V$ and $U(1)_A$ which are $\tau_3$ and 
$\One$ respectively. In the low density phase the rotation generated by
$\One$ gives the pion: $\bar\chi\varepsilon\chi$, while the rotation
generated by $\tau_3$ leaves the condensate invariant. Rotation of the diquark condensate by $\One$ gives a pseudoscalar diquark: $\chi^{tr}\tau_2\varepsilon\chi+
\bar\chi\tau_2\varepsilon\bar\chi^{tr}$,which is pseudo-Goldstone
for $\mu\neq 0$ and rotation by $\tau_3$ gives a scalar diquark: $\chi^{tr}\tau_2\chi-\bar\chi\tau_2\bar\chi^{tr}$ , which remains
an exact Goldstone mode even for $m\neq 0$. The fact that the low and high density regimes are characterised by different numbers of massless modes indicates that they should be separated by a true phase transition at some $\mu=\mu_c$. 
In general we expect $\mu_c=m_{lb}/N_{c}$ where $m_{lb}$ is the mass of the lightest baryon and $N_c$ is the number of colours. The $SU(2)$ theory is a special case because we expect $\mu_{c}=m_{\pi}/2$ i.e. there is a Goldstone baryon in the theory due to the $SU(2)$ symmetry which relates quark to anti-quark. In this respect the model is crucially different from QCD where we expect $\mu_{c}=m_{proton}/3$ yet find $\mu_{c}=m_{\pi}/2$ due to pathologies in the hybrid
Monte Carlo simulations for $\mu\neq 0$.

We can formally introduce a diquark source term into the fermionic sector
of the partition function as follows:
\be
Z_{\rm ferm.}=\int d\bar{\psi}d\psi \exp \left(\bar{\psi}M\psi +\psi J\psi
+\bar{\psi}\bar{J}\bar{\psi}\right)
\ee
where $M$ is the fermion matrix while
 $J$ and $\bar{J}$ are diquark and anti-diquark source
terms respectively. It is necessary to recast this equation in
order to perform the integration over the fermion fields. To do so we 
define a new {\bf 2 component Grassmann field}: 
$\phi\equiv \left({\begin{array}{c}\bar{\psi}\\
                                   \psi    \end{array}} \right)$
and then reconstruct $Z_{\rm ferm.}$ in terms of a ``double matrix''
\begin{equation}
Z_{\rm ferm.}=\int d\phi (\bar{\psi}\, \psi)\left(\begin{array}{cc}\bar{J}&\frac{M}{2}\\
                                           -\frac{M^{tr}}{2}&J\end{array}\right)    
\left(\begin{array}{c}\bar{\psi}\\ \psi \end{array}\right)
={\left\{{\det\left( \begin{array}{cc} j\tau_2 & \frac{M}{2}\\ -\frac{M^{tr}}
{2} & 
j\tau_2 \end{array}\right)}\right\}}^{1/2}
\end{equation}
where we have used 
$
\int d\phi \exp (\phi A \phi)=\sqrt{\det A}\equiv {\rm Pfaffian}(A).
$
Note that we require an antisymmetric matrix therefore
we chose $J=\bar{J}=j\tau_2$ with $j$ a real number. 

Making use of the property $\ln \det A =Tr \ln A$, on
the double matrix leads to the {\sl lattice diquark condensate}: 
\begin{equation}
\langle\psi\psi \rangle=\lim_{j\rightarrow 0}
\frac{\partial \ln Z}{\partial j}
=\lim_{j\rightarrow 0}\frac{1}{2}\Bigg<{Tr\left\{ {
\left(\begin{array}{cc} j\tau_2 & \frac{M}{2}\\ -\frac{M^{tr}}{2} & 
j\tau_2 \end{array}\right)}^{-1}\left(\begin{array}{cc}
0&0\\ 0 & 
\tau_2  \end{array}\right)\right\}}\Bigg>
\end{equation}

\begin{figure}[t]
\epsfaxhax{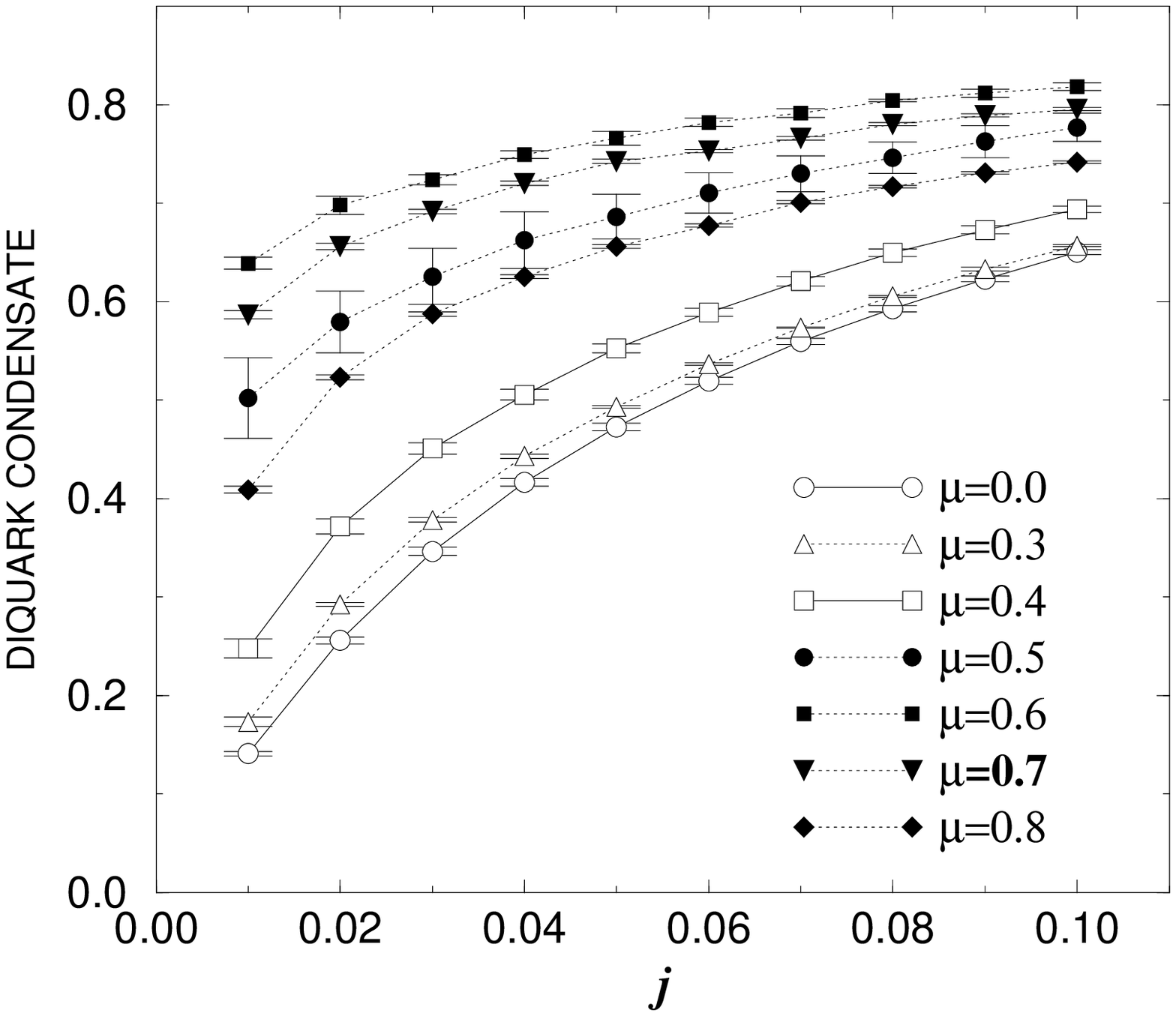}{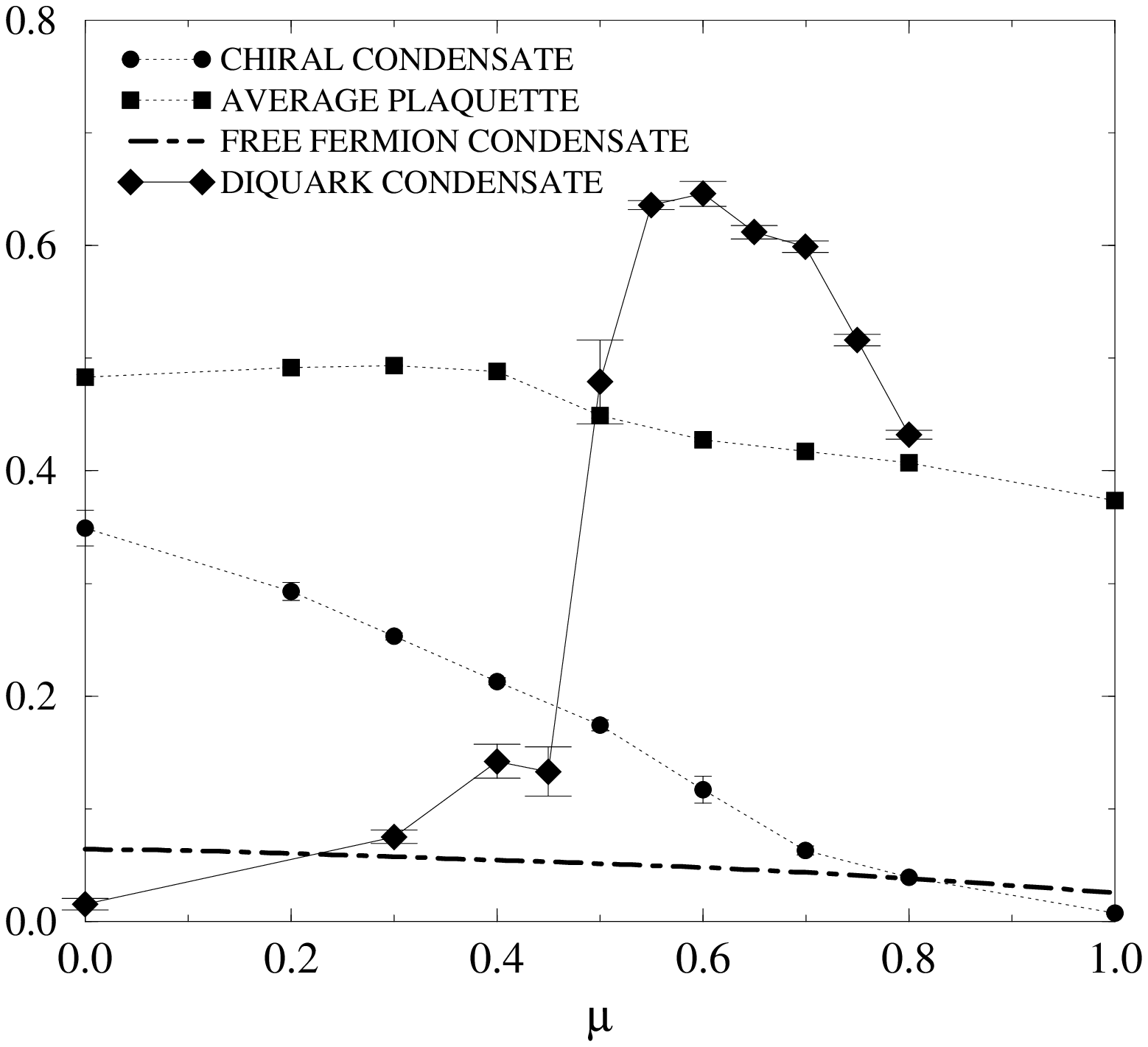}
\vskip -0.5cm
\caption{\hskip 4.1cm Figure 2:}
\vskip -0.5cm
\setcounter{figure}{2}
\end{figure}

In Fig. 1 we show the diquark condensate as a function of $j$ and
$\mu$. Polynomial fits to these curves were used to obtain the $j\to 0$
limit plotted alongside the standard observables in Fig. 2. The sharp 
rise in the diquark condensate at $\mu\simeq0.4$ coincides with the
downward jump in the average plaquette while the chiral condensate
decreases smoothly as $\mu$ is increased.


Measurements of the susceptibilities (Figs. 3,4) are strongly suggestive of
a phase transition at $\mu_c=0.4$. This is consistent with the 
conclusion from the observables in Fig. 2.
We shall refer to $\mu<m_{\pi}/2\simeq 0.4$ as the low density phase
and $\mu>m_{\pi}/2$ as the high density phase.
We find, as expected, that at $\mu=0$ the  scalar diquark is degenerate with the pion while the 
 pseudoscalar diquark is degenerate with the scalar meson.
The  pseudoscalar diquark which is heavy in the low density phase
 becomes light in the high density phase whereas the
pion which is a pseudo-Goldstone in
the low density phase becomes heavy in dense (chirally symmetric) phase. 
As seen in Fig. 4 susceptibilities of scalar and pseudoscalar diquarks 
become very large in the dense phase. In the dense phase we observe 
that the scalar diquark and the pseudoscalar diquark are light states
while the connected contributions to the scalar imply that it is
comparatively heavy. In our symmetries analysis we identified the scalar 
diquark as an exactly massless Goldstone mode in the dense phase while we 
expected
the pseudoscalar diquark to be a pseudo-Goldstone mode for $m\neq0$. The 
scalar meson is not expected to be a Goldstone mode in the high density phase.
Although the pion and scalar meson appear to be approximately degenerate in 
the high density phase this degeneracy may be broken when the disconnected 
contributions to the scalar are taken into account.

\begin{figure}[t]
\epsfaxhax{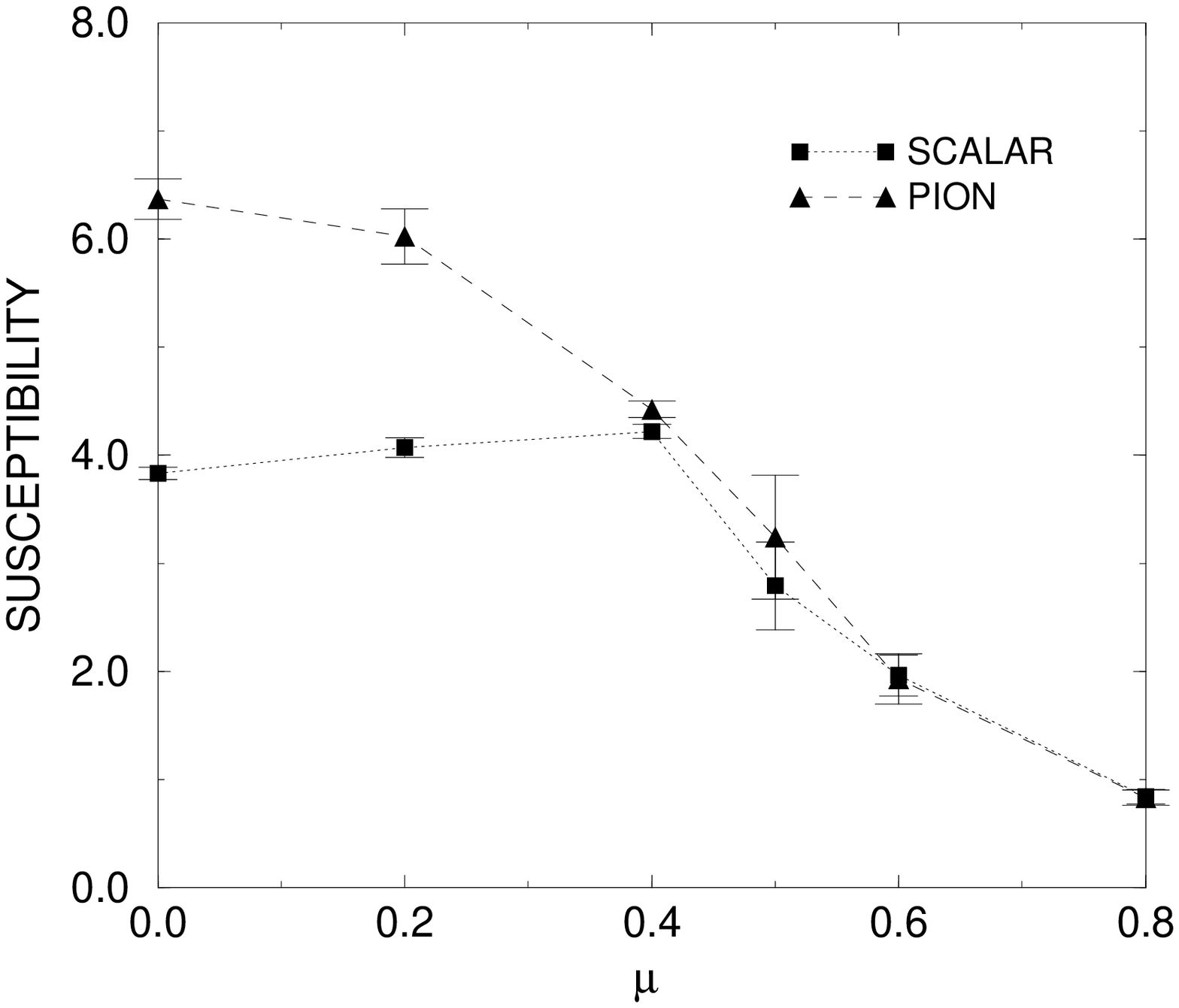}{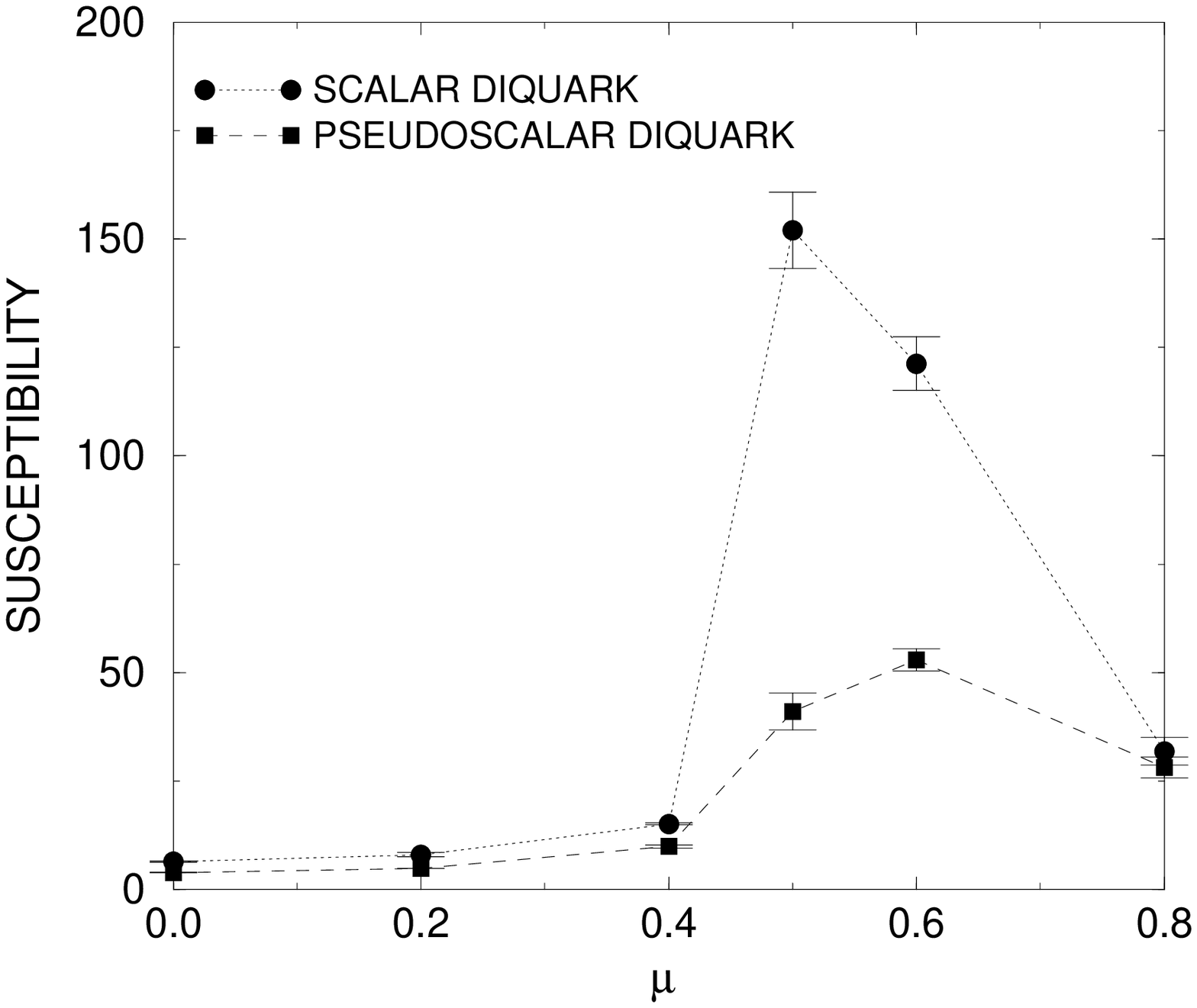}
\vskip -0.5cm
\caption{\hskip 4.1cm Figure 4:}
\vskip -0.5cm
\end{figure}


\end{document}